\documentclass[aps,prl,showpacs,twocolumn,preprintnumbers,superscriptaddress,amsmath,amssymb,amsfonts,floatfix]{revtex4-1}

\usepackage{graphicx,color,rotating,textcomp}
\usepackage{dcolumn}
\usepackage{bm}

\begin{document}

\title{Destroyed quantum Hall effect in graphene with [0001] tilt grain boundaries}

\author{Anders Bergvall}
\affiliation{Department of Microtechnology and Nanoscience - MC2,
Chalmers University of Technology, SE-412 96 G\"oteborg, Sweden}

\author{Johan M. Carlsson}
\affiliation{Accelrys Ltd, Cambridge Science Park 334, Cambridge CB4 OWN, United Kingdom}

\author{Tomas~L\"ofwander}
\affiliation{Department of Microtechnology and Nanoscience - MC2,
Chalmers University of Technology, SE-412 96 G\"oteborg, Sweden}

\date{\today}

\begin{abstract}
The reason why the half-integer quantum Hall effect (QHE) is suppressed in graphene grown by
chemical vapor deposition (CVD) is unclear. We propose that it might be connected to extended defects
in the material and present results for the quantum Hall effect in graphene with [0001] tilt grain boundaries
connecting opposite sides of Hall bar devices.
Such grain boundaries contain 5-7 ring complexes that host defect states that
hybridize to form bands with varying degree of metallicity depending on grain boundary defect density.
In a magnetic field, edge states on opposite sides of the Hall bar
can be connected by the defect states along the grain boundary. This destroys 
Hall resistance quantization and leads to non-zero longitudinal resistance.
Anderson disorder can partly recover quantization, where current instead flows along returning paths
along the grain boundary depending on defect density in the grain boundary and on disorder strength.
Since grain sizes in graphene made by chemical vapor deposition are usually small,
this may help explain why the quantum Hall effect is usually poorly developed in devices made of this material.
\end{abstract}


\maketitle

The half-integer quantum Hall effect (QHE) \cite{NovoselovNature2005,ZhangNature2005} in monolayer graphene
grown on silicon-carbide substrates has been observed to
metrological accuracy \cite{TzalNatureNano2010,PoirierCRP2011,JanssenMet2012}.
Very high breakdown currents have been recorded, and quantization remains accurate also
at elevated temperatures. This material may therefore be the next choice for an improved resistance standard.
On the other hand, QHE plateaux have not been measured to the same level of accuracy on Hall bars
made of graphene grown by chemical vapor deposition (CVD) \cite{NamAPL2013,Lafont2014}.
The reason for this disparity is unclear, but it may be due to extrinsic
effects, such as defects and inhomogeneity introduced in the process of graphene transfer from substrates
used in the growth to other substrates used for devices, or due to defects in the material itself,
such as grain boundaries that usually are found in graphene made by CVD \cite{BanhartACSNano2011}.

In a recent experiment  \cite{Lafont2014}, it was indeed argued that grain boundaries may
be the source of reduced quantization in devices made of CVD graphene.
A clear theoretical picture of how the QHE is destroyed in graphene with grain boundaries is however still lacking.
One particular and very special type of grain boundary has been considered theoretically 
in the literature before \cite{YaoPRB2013,Lafont2014}.
The grain boundary consists of a perfect row of 5-8-5 ring complexes
that separates two perfect armchair ribbons oriented along the same axis.
To join the armchair ribbons to the grain boundary, the ribbons are cut at $90^{\circ}$ to their armchair edges
so that perfect zigzag edges are formed. These zigzag edges can be attached to the grain boundary.
In a magnetic field, a picture appears of current flowing along an armchair edge in the ribbon and along
a zigzag edge along the grain boundary over to the opposite edge of the ribbon where the current
can flow back in the opposite direction. This special type of grain boundary is not the only or typical grain boundary in
graphene \cite{BanhartACSNano2011,HuangNature2011,KimACSNano2011,AnACSNano2011,NemesCarbon2013}
and a more extensive investigation of other grain boundaries is called for.

\begin{figure}[b]
\includegraphics[width=0.7\columnwidth]{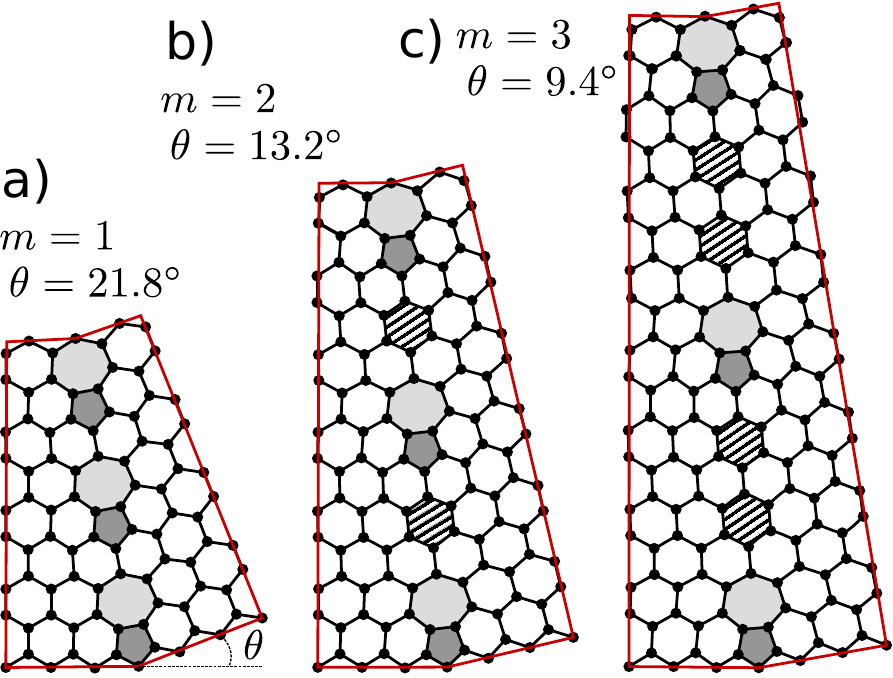}
\caption{The first three in a series of grain boundaries with decreasing misiorientation angle $\theta$.
The grain boundaries contain 5-7 ring defects on a line with an increasing number of hexagons (hatched)
between neighboring defects.}
\label{fig_grains}
\end{figure}

Here, we report an investigation of the influence of [0001] tilt grain boundaries on the QHE in graphene.
We show by numerical simulation that electronic states at dislocation cores (5-7 ring complexes),
which form several 1D metallic bands along the grain boundary, can in a strong externally applied magnetic field
connect two edge states on opposite sides of the Hall bar and thereby destroy quantization.
The resulting conductance fluctuations depend on the number of dislocation cores in the grain boundary,
which is related to the grain boundary tilt angle and its physical length.
This is similar to the situation for graphene grown on silicon-carbide substrates, where it has been shown
by numerical simulation \cite{LofwanderPRB2013} and experiment \cite{YagerNL2013}
that bilayer stripe defects connecting Hall bar edges destroy the QHE in that material. Only in samples
without such large bilayer defects one may expect the precise quantization reported in \cite{JanssenMet2012}.

The grain boundary models were constructed by the coincidence site lattice (CSL) theory according to the method presented in Ref.~\cite{CarlssonPRB2011}.  We focus on a series of grain boundaries for which the Burger's vector is one lattice vector long ($n_d=1$ series in Ref.(~\cite{CarlssonPRB2011}), meaning that we limit ourselves to a class of [0001] tilt grain boundaries with one 5-7 ring complex per grain boundary unit cell, but an increasing number of hexagons per unit cell for
decreasing grain boundary angles. This means that the distance between 5-7 ring defects
increases by one hexagon as we move along the $n_d$=1 series towards smaller tilt angles, see Fig.~\ref{fig_grains}.
We consider the first six grain boundaries in the series, which we enumerate by $m=1,2,...,6$, for which the grain
boundary angles are $\theta=21.8^{\circ}$, $13.2^{\circ}$, $9.4^{\circ}$, $7.3^{\circ}$, $6.0^{\circ}$, and $5.1^{\circ}$. The geometry was optimized by a force field calculation using the Dreiding Force Field as implemented in the Forcite module in Materials Studio ~\cite{MS70}. We aim to model graphene grown on a substrate, so the graphene sheet was kept flat even at the grain boundaries by fixing the atomic relaxation perpendicular to the sheet during the relaxation.
Using the atomic positions of the relaxed grain boundary, we create a two-terminal zigzag nanoribbon of varying width and study electron transport in a magnetic field.

In the transport simulations, the system is modeled by a tight-binding Hamiltonian
\begin{equation}
H  = \sum_{i}\epsilon_ic_i^{\dagger}c_i +  \sum_{ij} t_{ij}c_i^{\dagger}c_j.
\label{eq_hamiltonian}
\end{equation}
The onsite energies are either put to zero, as in defect-free
graphene, or we include Andersson disorder by putting $\epsilon_i$ to
random numbers uniformly distributed between $[-W/2,W/2]$, where $W$ is
the disorder strength.
The atomic positions determine the hopping elements $t_{ij}$
through an approximative formula for $\pi$-orbital overlap at
different carbon sites $j$ and $i$ separated by ${\bf R}_j-{\bf  R}_i={\bf r}=(x,y)^T$,
\begin{equation}
t_{ij} = t({\bf r}) = -\gamma_0 \frac{x^2+y^2}{r^2}e^{-\lambda(|{\bf r}|-a_{cc})},
\label{eq_hopping}
\end{equation}
where $\gamma_0$ is the nearest neighbor hopping parameter,
$a_{cc}$ is the carbon-carbon distance,
and the exponent is $\lambda\approx 3/a_{cc}$. The formula in
Eq.~(\ref{eq_hopping}) is applied for atomic distances $r=|{\bf r}|$
reaching a cut-off $R_c$, beyond which $t_{ij}=0$. For the present problem,
a good description is obtained for small $R_c\gtrsim a_{cc}-2a_{cc}$.
Going beyond the nearest neighbor approximation means that the Dirac point
in the bandstructure obtained from the model in Eq.~(\ref{eq_hamiltonian})
is shifted from zero to a higher energy $E_{Dirac}\approx 0.3\gamma_0$.
The applied magnetic field enters the hopping parameters through a standard Peierls substitution,
$t_{ij} \rightarrow \exp\left[ie\int_{{\bf R}_j}^{{\bf R}_i} {\bf A}\cdot {\bf dl/\hbar}\right]t_{ij}$.
%
%
The magnetic field enters naturally as an applied magnetic flux per hexagon (hexagon area of
perfect graphene) in units of the flux quantum $\Phi_0=h/e$.
The magnetic field defines the magnetic length through $\ell_B=\sqrt{\hbar/(|e|B)}$,
which in all simulations is smaller than the ribbon width, which means that the spectrum
is dominated by Landau levels $E_n-E_{Dirac}=\sqrt{2n}\hbar v_f/\ell_B=\sqrt{n}\omega_c$,
where $n\ge 0$ is an integer, and $v_f$ if the electron velocity of graphene in the absence of magnetic field.
In the ribbon geometry, Landau level bands acquire dispersive parts corresponding to the well
known edge states which carry the current in the quantum Hall regime
(see for instance Fig.~2 in Ref.~\cite{LofwanderPRB2013} for an illustration).

\begin{figure}[t]
\includegraphics[width=0.8\columnwidth]{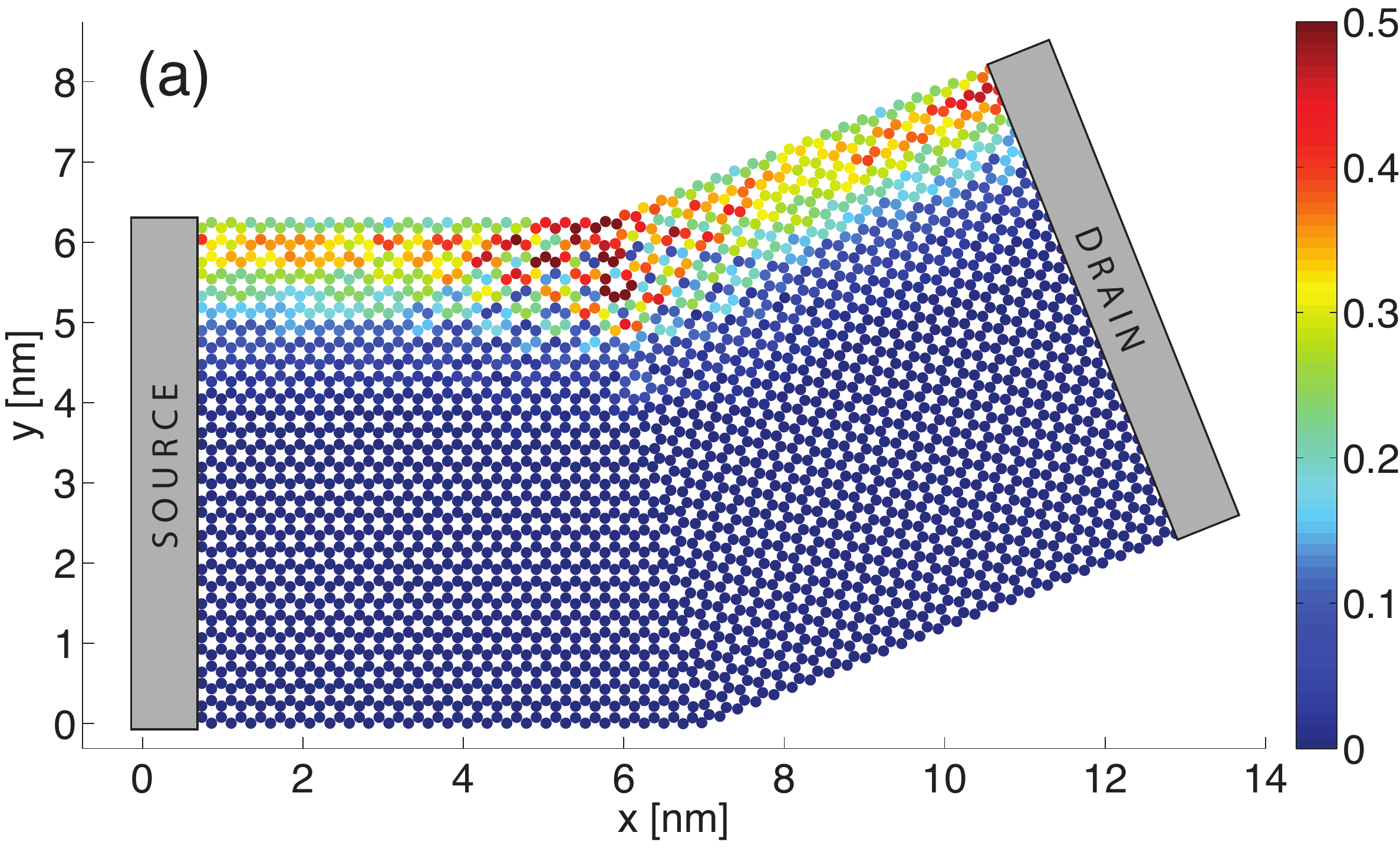}
\includegraphics[width=0.9\columnwidth]{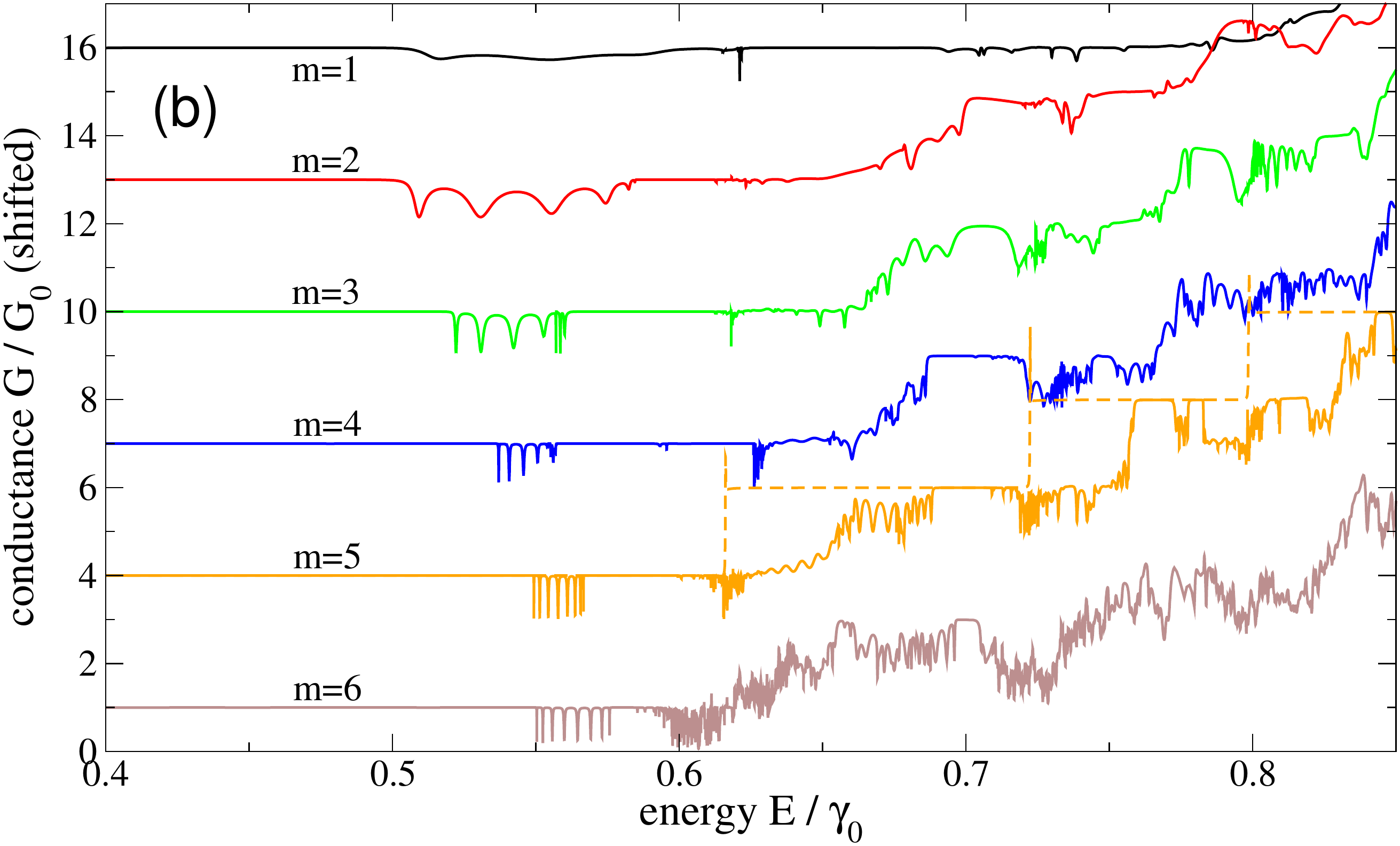}
\caption{(a) A nanoribbon with a $m=1$ grain boundary. Current flows along the upper edge
(color scale denotes the absolute value of the current in units of $G_0V$,
where $V$ is the applied voltage between source and drain)
at energy $E=0.45\gamma_0$ for which the conductance is quantized, $G=G_0$.
(b) Conductance at zero temperature for nanoribbons with one grain boundary
including $10$ defects as function of energy. The different curves (shifted by $3G_0$ for clarity)
correspond to different grain boundary angles and therefore also different ribbon widths
since the defect density varies with grain boundary angle.
The dashed orange line is the quantized conductance of an ideal ribbon without grain boundary shifted by $3G_0$.
The magnetic field corresponds to a flux $\Phi=0.01\Phi_0$ per hexagon in all cases.
In (a) the Anderson disorder strength is $W=0.25\gamma_0$, while it is $W=0$ in (b).}
\label{fig_system}
\end{figure}

In a magnetic field the two-terminal conductance equals the transverse
conductivity $\sigma_{xy}$ in a Hall bar geometry when the contacts are perfect (as
is the case in these simulations). The two-terminal conductance at zero temperature $G=G_0T(E)$
is given in terms of the conductance quantum $G_0=2e^2/h$
and the linear response transmission function $T(E)$.
The latter is computed through the retarded propagator of the system
$G^R_{ij}(E)$ and self-energies of the lead surfaces
$\Sigma_{\ell}^R(E)$, which remain after the leads have been
eliminated in favor of the system in a standard way,
see for instance Ref.~\cite{DattaBook}. The leads are
enumerated by the index $\ell$ ($\ell=1$ and $2$ for source and
drain). The formula for the transmission is then
\begin{equation}
T(E) = \mbox{Tr}\left[ \Gamma_1(E) G^R_{12}(E) \Gamma_2(E) G^A_{21}(E) \right]
\end{equation}
where $\Gamma_{\ell}=i[\Sigma_{\ell}^R -(\Sigma_{\ell}^R)^{\dagger}]$,
and $G^R_{12}$ symbolises the propagator between leads $1$ and $2$.
The advanced propagator $G^A_{21}(E)$ is the hermitian conjugate of
the retarded propagator and the trace is over the surface sites.

For local current flow patterns we need the lesser Green's function
$G^<$. In the absence of electron correlations, the lesser Green's
function is reduced to the form
\begin{equation}
G_{ij}^<(E) = \sum_{\ell} f_{\ell}(E)
\sum_{c\tilde c} G_{ic}^R(E)
\left[ \Gamma_{\ell}(E) \right]_{c\tilde c}
G_{\tilde cj}^A(E),
\label{Glesser}
\end{equation}
which involves the distribution functions of the leads $f_{\ell}(E)$.
Surface sites of the leads are labeled by $c$ and $\tilde c$. Local
charge current flow in the device (bond current between sites $i$ and
$j$) is then written as
\begin{equation}
I_{ij} = e\int_{-\infty}^{\infty}
\left[ t_{ij}G_{ji}^<(E)-t_{ji}G_{ij}^<(E) \right]dE
=\int_{\infty}^{\infty}I_{ij}(E)dE.
\label{bondcurrent}
\end{equation}
The retarded Green's function $G _{ij}^R(E)$ of the system is computed
numerically through our own implementation of a recently developed
recursive algorithm \cite{Waintal} within which sites are added one
by one. Below we present the spectral current flow pattern
$I_{ij}(E)$ assuming zero temperature and current injection from one contact.

In Fig.~\ref{fig_system}(a) we display a narrow nanoribbon with an $m=1$
grain boundary with ten 5-7 ring defect complexes.
In a strong magnetic field, the current enters for instance from the top left corner and flows along the edge.
Without scattering against defects, as in Fig.~\ref{fig_system}(a), the current
reaches the right contact and is absorbed. Depending on the electron
density, i.e. the location of the Fermi energy, we have different
number of Landau levels occupied and the corresponding number of edge states.
Each edge state carries a unit of conductance including spin degeneracy, $G_0$,
since we neglect the Zeeman effect. The $n=0$ level is special
in that valley degeneracy in the bulk is broken in the finite size
ribbon for the dispersive (edge state) parts of the spectrum.
Higher Landau levels have valley degenerate edge states. The conductance sequence
at zero temperature is therefore $G=(2n+1)G_0$ for
energies above the Dirac point (electron doping), and the same sequence for increasing hole doping.
This sequence of plateaux is illustrated in Fig.~\ref{fig_system}(b) by the orange dashed line
for electron doping (energies $E>E_{Dirac}\approx 0.3\gamma_0$).

\begin{figure}[t]
\includegraphics[width=0.9\columnwidth]{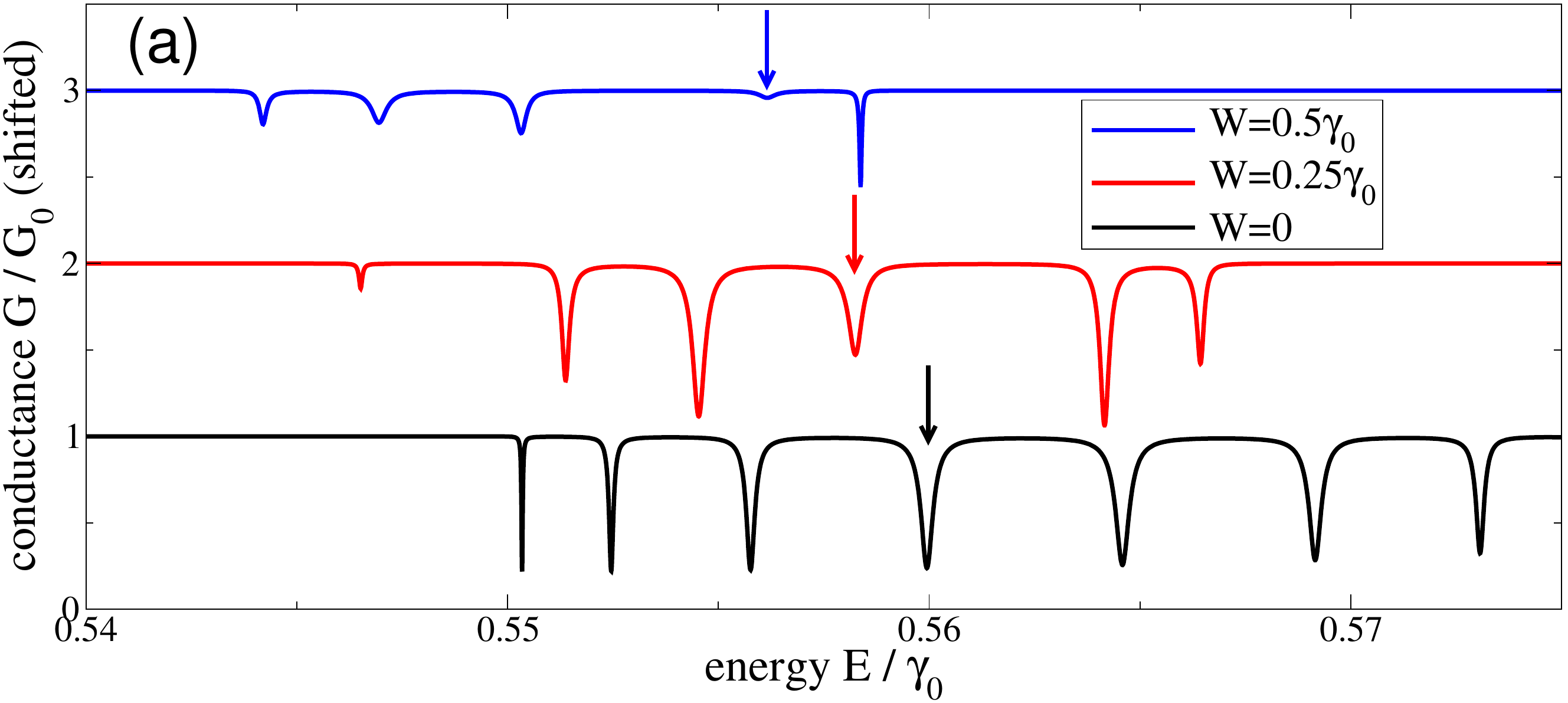}
\includegraphics[width=\columnwidth]{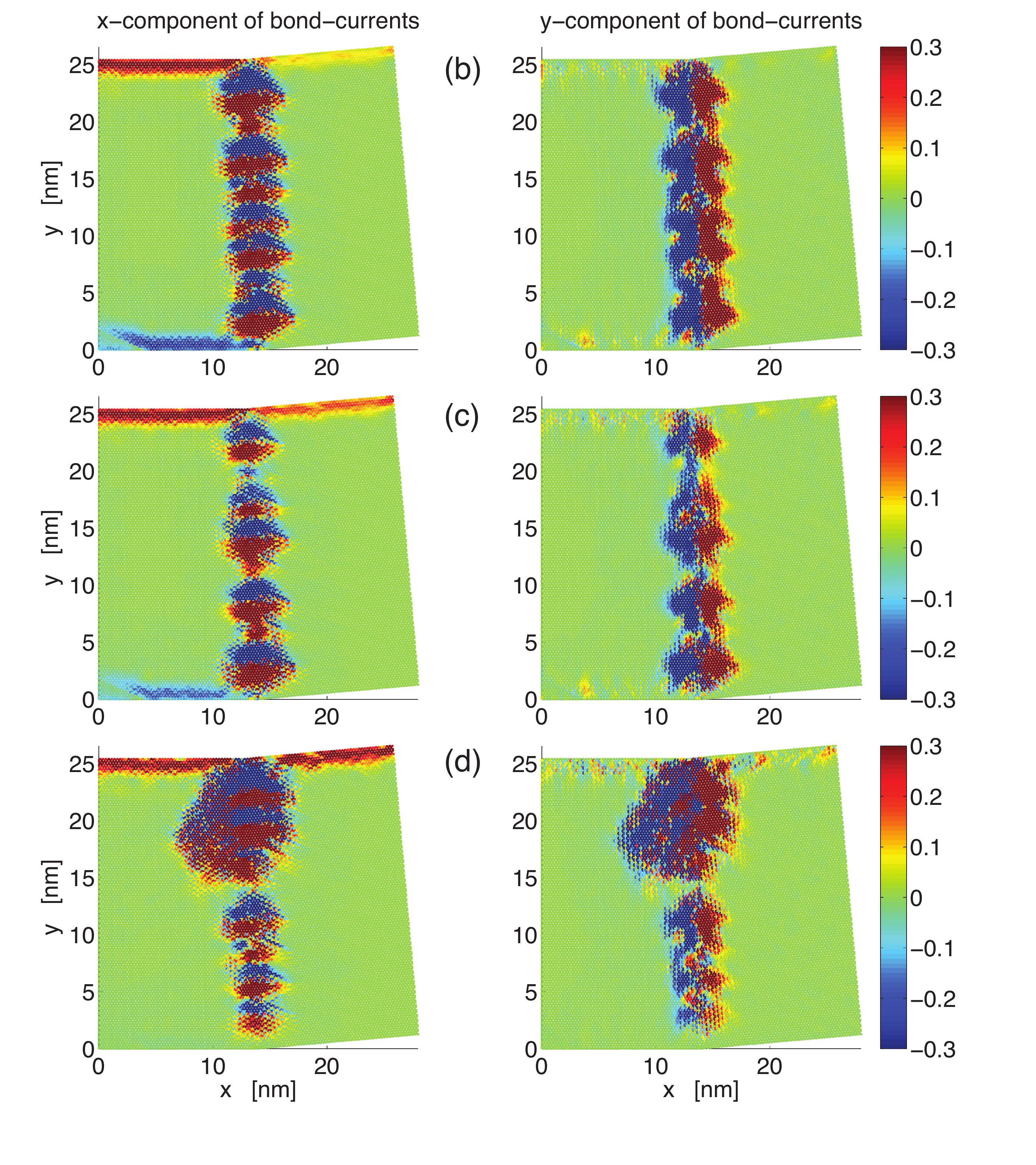}
\caption{(a) Evolution of the conductance in Fig.~\ref{fig_system}(b) for the $m=6$ grain boundary
with increasing Anderson disorder strength $W$ (other model parameters are held fixed).
(b)-(d) Examples of local current flow patterns near the grain boundary for
(b) $W=0$, (c) $W=0.25\gamma_0$, and (d) $W=0.5\gamma_0$.
The energies where these current flow patterns appear are marked in (a) by vertical arrows.
Disorder are distributed randomly across the whole displayed systems, with ideal source and drain contacts
attached to the left and right ribbons.}
\label{fig_frames}
\end{figure}

Let us now discuss the influence of the grain boundaries.
It is well known that at each 5-7 defect, there are defect states. They may
hybridize along the grain boundary, which then becomes metallic, as has been discussed in
several papers \cite{Lahiri,YazyevSSC2012,IhnaPRB2013,GargiuloNanoLett2014}. In a
magnetic field, with the grain boundary connecting the upper and lower
edges, the question arises what will happen with the edge state current flow.
In Fig.~\ref{fig_system}(b) we present the conductance as function of energy, corresponding
to the Fermi energy at zero temperature (which can be controlled by a back gate in
a real device). For large grain boundary misorientation angles, $m=1$ and $2$ in Fig.~\ref{fig_system}(b), 
the metallicity of the grain boundary is apparent as the plateaux are not quantised.
Edge states at opposite edges (which carry current in opposite directions) are connected
by the metallic grain boundary which causes partial reflection of current and destruction of the plateaux.
For lower grain boundary angles, the 5-7 defects are further apart and hybridization is less effective.
In a magnetic field, a more pronounced reflection resonance pattern then develops, see for instance the $m=6$ 
grain boundary conductance curve in Fig.~\ref{fig_system}(b) (brown line).
This resonance behavior is particularly clear for the $n=0$ plateau, where a comb of resonances
are well defined near $E\approx 0.56\gamma_0$. The number of resonances (size of the comb)
depends on the number of defects along the grain boundary. In Fig.~\ref{fig_system}(b) the ribbon width
is varying so that the grain boundaries always hold ten 5-7 defects. The comb is present for all grain boundary
angles, but the enhanced hybridization and enhanced
metallicity for grain boundaries with more dense defect densities is clear in Fig.~\ref{fig_system}(b)
when comparing the combs' shapes for grain boundaries with decreasing $m$. For higher energies, many more resonances
appear and quantization for higher Landau levels is completely destroyed.

We can gain additional insight into the nature of the conductance
fluctuations by looking at the local current flow patterns. In Fig.~\ref{fig_frames}(a) we
display a zoom of the resonance comb on the $n=0$ plateau for the $m=6$ grain boundary (black curve for $W=0$).
For the energy indicated by the vertical black arrow, we present the local current flow pattern in Fig.~\ref{fig_frames}(b).
The current flows in a circular fashion around each 5-7 ring defect and at the same time displays an envelope
pattern across the entire grain boundary from the upper to the lower ribbon edges. The envelope contains a varying number
of nodes for the different resonances in the comb. Each resonance, therefore, corresponds to a particular hybridization
of the defect states along the grain boundary.

In Fig.~\ref{fig_frames}(a), we study the evolution of the resonance comb with increasing Anderson disorder strength $W$.
For $W=0.25\gamma_0$ ($\sim0.68$ eV for $\gamma_0\approx 2.7$ eV), the resonances are shifted and some are
weakened. For $W=0.5\gamma_0$ this effect is more pronounced, and at higher $W$ resonances disappear.
The current flow patterns for the conductance dips marked
by the red arrow ($W=0.25\gamma_0$) and the blue arrow ($W=0.5\gamma_0$)
are shown in Fig.~\ref{fig_frames}(c) and (d), respectively. For increasing disorder strength, the current flows chaotically down
the grain boundary but eventually [for strong $W$, Fig.~\ref{fig_frames}(d)]
a situation resembling localization along the grain boundary appears and
quantization is improved. This picture of localization agrees with the recent results in Ref.~\cite{Lafont2014} for the special
5-8-5 line defect mentioned in the introduction.

\begin{figure}[t]
\includegraphics[width=0.8\columnwidth]{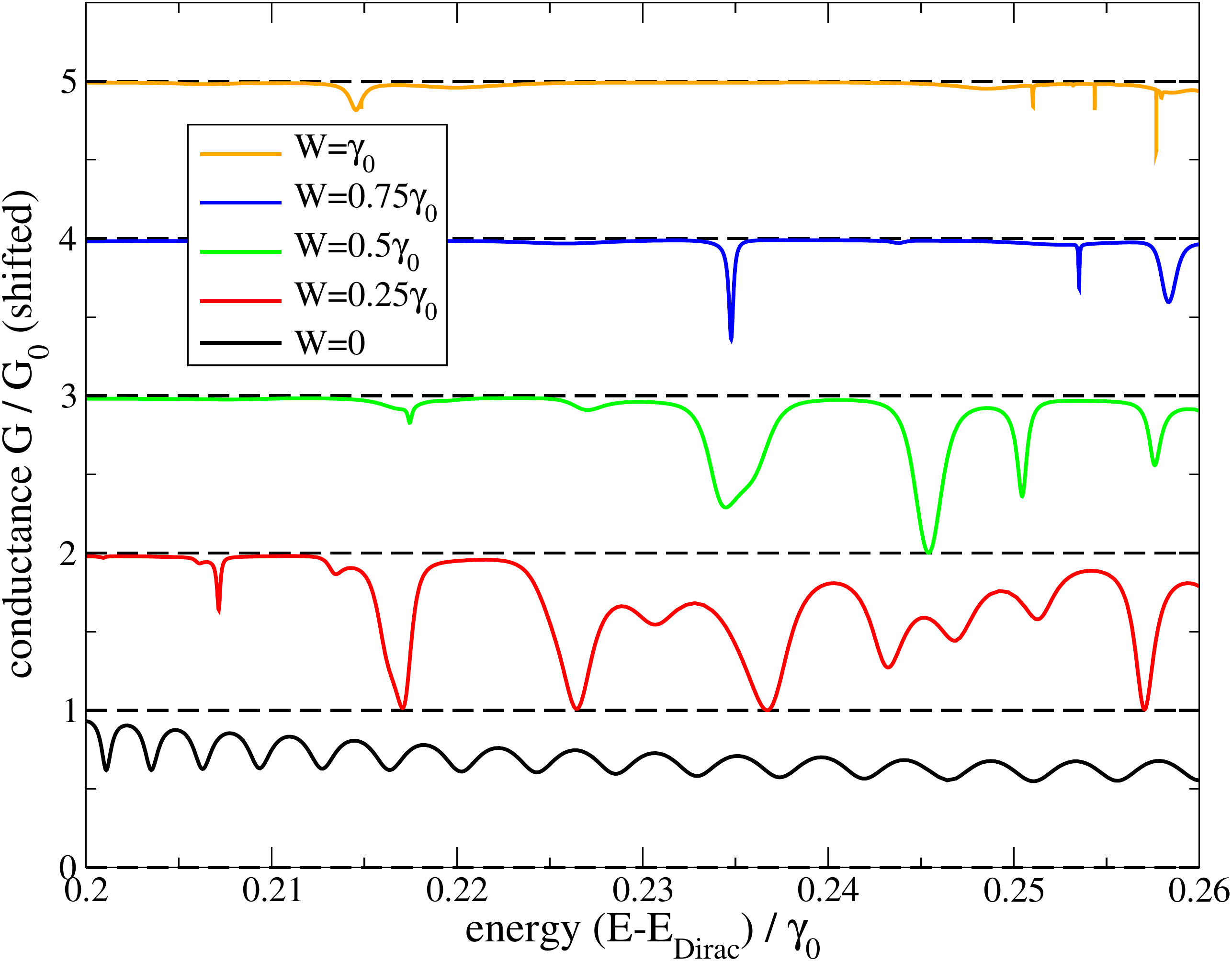}
\caption{Conductance on the lowest plateau at zero temperature for a 50 nm wide nanoribbon with
a $m=1$ grain boundary for varying Anderson impurity disorder strength.
The different curves are shifted by $G_0$ for clarity and 
the magnetic field corresponds to a flux $\Phi=0.01\Phi_0$ per hexagon.}
\label{fig_S7}
\end{figure}

In Fig.~\ref{fig_S7}, we study the effect of Andersson disorder for a $50$ nn wide ribbon with a $m=1$ grain boundary.
This grain boundary has the most dense defect density and display for $W=0$ non-quantized conductance reflecting
the metallicity of the grain boundary (black wavy curve in the figure). Introducing weak Andersson disorder $W=0.25\gamma_0$,
leads to development of more sharp resonances (dips with zero conductance; red curve in the figure).
For larger $W$, also these resonances disappear. The sensitivity of resonances to disorder depends on the grain boundary
angle. Smaller grain boundary angles correspond to less defect density, sharper resonances, and higher sensitivity to disorder.
For system sizes that we have considered, higher plateaux are however always destroyed, while the first plateau is more robust.

In conclusion, we have studied the influence of [0001] tilt grain boundaries on the quantum Hall effect in
granular graphene. We find that electronic states formed at dislocation cores (5-7 complexes) in the grain boundary
form metallic bands that in a magnetic field can short circuit counter propagating edge states on
opposite sides of the Hall bar. The QHE is thereby destroyed. Depending on the defect density along the grain
boundary, weak Andersson disorder can lead to recovery of at least the $n=0$ plateau.
This indicates that the fact that the QHE so far has not been observed
to metrological accuracy in CVD graphene could be due to the granularity of this material.

\acknowledgements
We acknowledge financial support from the EU through FP7 STREP ConceptGraphene,
the Swedish Foundation for Strategic Research, and the Knut and Alice Wallenberg foundation.

\end{document}